\documentclass[english,aps,pra,showpacs,twocolumn,floatfix,superscriptaddress]{revtex4-1}
\usepackage{mathptmx}
\usepackage[T1]{fontenc}
\usepackage[latin9]{inputenc}
\usepackage{color}
\usepackage{verbatim}
\usepackage{bm}
\usepackage{amsmath}
\usepackage{graphicx}
\usepackage{amssymb}

\makeatletter
\@ifundefined{textcolor}{}
{%
 \definecolor{BLACK}{gray}{0}
 \definecolor{WHITE}{gray}{1}
 \definecolor{RED}{rgb}{1,0,0}
 \definecolor{GREEN}{rgb}{0,1,0}
 \definecolor{BLUE}{rgb}{0,0,1}
 \definecolor{CYAN}{cmyk}{1,0,0,0}
 \definecolor{MAGENTA}{cmyk}{0,1,0,0}
 \definecolor{YELLOW}{cmyk}{0,0,1,0}
 }

\makeatother

\usepackage{babel}

\begin{document}

\title{Photonic band-gap engineering for volume plasmon polaritons \\
in multiscale multilayer hyperbolic metamaterials}

\author{Sergei V. Zhukovsky}

\email{sezh@fotonik.dtu.dk}

\affiliation{DTU Fotonik -- Department of Photonics Engineering, Technical University
of Denmark, {\O}rsteds Pl. 343, DK-2800 Kgs.~Lyngby, Denmark}

\affiliation{National Research University of Information Technology, Mechanics
and Optics, Kronverksky pr. 49, St. Petersburg, 197101, Russia}

\author{Alexey A. Orlov}

\affiliation{National Research University of Information Technology, Mechanics
and Optics, Kronverksky pr. 49, St. Petersburg, 197101, Russia}

\author{Viktoriia E. Babicheva}

\affiliation{DTU Fotonik -- Department of Photonics Engineering, Technical University
of Denmark, {\O}rsteds Pl. 343, DK-2800 Kgs.~Lyngby, Denmark}

\affiliation{National Research University of Information Technology, Mechanics
and Optics, Kronverksky pr. 49, St. Petersburg, 197101, Russia}

\affiliation{Birck Nanotechnology Center, Purdue University, 1205 West State Street,
West Lafayette, IN, 47907-2057 USA}

\author{Andrei V. Lavrinenko}

\affiliation{DTU Fotonik -- Department of Photonics Engineering, Technical University
of Denmark, {\O}rsteds Pl. 343, DK-2800 Kgs.~Lyngby, Denmark}

\author{J. E. Sipe }

\affiliation{Department of Physics and Institute for Optical Sciences, University
of Toronto, 60 St.~George Street, Toronto, Ontario, M5S 1A7, Canada.}
\begin{abstract}
We theoretically study the propagation of large-wavevector waves (volume
plasmon polaritons) in multilayer hyperbolic metamaterials with two
levels of structuring. We show that when the parameters of a subwavelength
metal-dielectric multilayer ({}``substructure'') are modulated ({}``superstructured'')
on a larger, wavelength scale, the propagation of volume plasmon polaritons
in the resulting \emph{multiscale hyperbolic metamaterials} is subject
to photonic band gap phenomena. A great degree of control over such
plasmons can be exerted by varying the superstructure geometry. When
this geometry is periodic, stop bands due to Bragg reflection form
within the volume plasmonic band. When a cavity layer is introduced
in an otherwise periodic superstructure, resonance peaks of the Fabry-P{\'e}rot
nature are present within the stop bands. More complicated superstructure
geometries are also considered. For example, fractal Cantor-like multiscale
metamaterials are found to exhibit characteristic self-similar spectral
signatures in the volume plasmonic band. Multiscale hyperbolic metamaterials
are shown to be a promising platform for large-wavevector bulk plasmonic
waves, whether they are considered for use as a new kind of information
carrier or for far-field subwavelength imaging.
\end{abstract}


\maketitle
\global\long\def\eff{\text{eff}}

\section{Introduction\label{sec:INTRO}}

\begin{figure}[b]
\begin{centering}
\includegraphics[width=1\columnwidth]{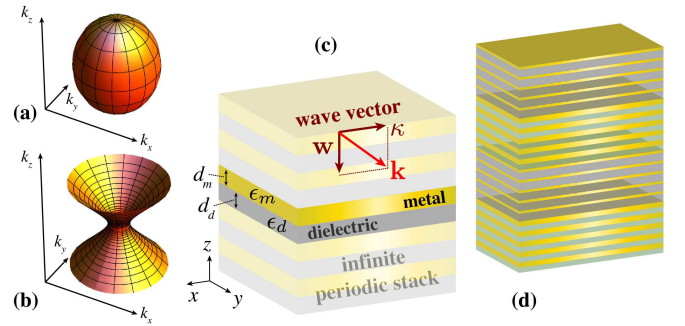}
\par\end{centering}

\caption{(Color online) Dispersion relation of (a) a conventional anisotropic
medium ($\epsilon_{x,y,z}>0$) with ellipsoidal isofrequency surface
and (b) an indefinite medium ($\epsilon_{x,y}<0$ and $\epsilon_{z}>0$)
with hyperboloidal isofrequency surface. (c) Schematic of a periodic
metal-dielectric multilayer with wave vector decomposition $\mathbf{k}=w\hat{\mathbf{z}}+\kappa\hat{\mathbf{x}}$
used in Section~\ref{SEC:PROPAGATION} (schematic in that both $w$
and $\mathbf{k}$ can be complex). (d) Schematic example of a multiscale
metal-dielectric HMM which is the main object of studies in this paper.
\label{FIG:Schematic}}
\end{figure}

Metamaterials have attracted avid scientific interest over the last
decade because optical materials with properties rare or absent in
nature can be artificially engineered. Notable examples include media
with negative refraction \cite{0-PendryFocusIssue} or giant optical
activity \cite{1-Giant05}, and so-called indefinite media, which
exhibit hyperbolic dispersion relations \cite{2Smith03,2aSmith04,2bSmith07,2cSurfaceIndef}.
The latter are a special case of extreme anisotropy where components
of the diagonalized permittivity tensor have opposite signs (e.g.
$\varepsilon=\text{diag}(\epsilon_{x},\epsilon_{y},\epsilon_{z})$
with $\epsilon_{x}=\epsilon_{y}<0$, $\epsilon_{z}>0$ for uniaxial
anisotropic media). With the introduction of these opposite signs,
the dispersion relation, \begin{equation}
\omega^{2}/c^{2}=k_{x}^{2}/\epsilon_{z}+k_{y}^{2}/\epsilon_{z}+k_{z}^{2}/\epsilon_{x,y},\label{eq:disp_initial}\end{equation}
changes from one of a conventional elliptical form to one of an \textquotedbl{}exotic\textquotedbl{}
hyperbolic form (see Fig.~\ref{FIG:Schematic}a--b). 

In the idealization that such a dispersion relation holds for all
wavevectors, the isofrequency surface in the dispersion relation becomes
unbounded (Fig.~\ref{FIG:Schematic}b). As a result, waves with very
large wave vectors ($k^{2}\gg\epsilon_{x,y,z}\omega^{2}/c^{2}$),
which would normally be evanescent in any isotropic medium, can become
propagating. Information carried by these {}``high-$k$ modes''
with anomalously small wavelength $2\pi/k$ can be used for far-field
subwavelength imaging, as in the recent proposal of a hyperlens \cite{0HyperlensOE06}.
In addition, a multitude of high-$k$ modes greatly increases the
local photonic density of states (DOS) in the indefinite medium, bringing
about a variety of new physical effects including broadband spontaneous
emission enhancement \cite{0NarimanovAPB10,0NarimanovAPL12}, anomalous
heat transfer beyond the Stefan-Boltzmann limit \cite{NarimanovStefanBoltzmann},
and optical {}``tabletop simulation platform'' for space-time phenomena
such as metric signature transitions \cite{5NarimanovPRL10,0SmolyaninovJOSA}.

Practical realization of indefinite media has been achieved over the
past few years in the form of hyperbolic metamaterials (HMMs), which
are highly anisotropic, subwavelength metal-dielectric composites.
Two geometries of HMMs have been preferred so far due to their simplicity
in both modeling and fabrication. They are (i) metallic nanorod arrays
embedded in a dielectric host \cite{0NoginovAPL09,3NoginovOL10,6bSchillingAPL10}
and (ii) metal-dielectric multilayers \cite{0NarimanovAPB10,0NarimanovAPL12}
shown in Fig.~\ref{FIG:Schematic}c. In both structures, the behavior
expected for indefinite media was experimentally confirmed \cite{3NoginovOL10,0NarimanovAPB10},
opening up many areas of theoretical and experimental research (see
recent reviews \cite{ReviewJacob,ReviewKildishev} and references
therein). The finite size of the structure elements (rods or layers)
puts an upper limit on the wave vectors $k$ that still satisfy Eq.~\eqref{eq:disp_initial}
\cite{0NarimanovAPB10,YuryOL11,0ourOL11}. Still, the existence of
such waves in HMMs has been clearly demonstrated \cite{AvrPRB,LasPhotRev}.
Since these waves underlie the operation of a hyperlens and are crucial
to other exotic physical properties of HMMs, it is very important
to understand the physical nature of these waves and investigate the
possible means of controlling their excitation and propagation.

The metal-dielectric composition of HMMs naturally leads one to suspect
that high-$k$ modes are plasmonic in nature. Even though the exact
mechanism of their formation has been debated \cite{blFengOE05,blOrenOE11},
it is generally accepted that high-$k$ propagating waves must originate
from surface plasmon excitations at individual metal-dielectric boundaries
\cite{AvrPRB}. Because of this, names such as \emph{multilayer plasmons}
\cite{Schilling}, \emph{Bloch plasmon polaritons} \cite{AvrAPL,AvrPRB},
or \emph{volume plasmon polaritons} (VPPs) \cite{LasPhotRev} have
been used by various groups. In our recent work \cite{ourFocusOE13},
it was shown that VPPs originate from coupling between short-range
surface plasmon polaritons in the individual metal layers. 

In order to utilize the full potential of VPPs as subwavelength information
carriers for hyperlensing and other applications, it is necessary
to understand how these waves can be guided and otherwise manipulated.
A remarkable thing to observe is that they are bulk propagating waves,
so they should be subject to the photonic band-gap (PBG) effects similar
to all other kinds of propagating waves. For an idealized model of
homogeneous indefinite medium, a photonic structure can be imagined
by imposing a modulation of medium parameters (e.g., $\epsilon_{x,y}$
and $\epsilon_{z}$), with the PBG properties depending on the geometry
of that modulation. For example, a periodic stepwise modulation is
expected to act as a photonic crystal for VPPs. In a realistic multilayer
HMM, one can similarly envisage adding a wavelength-scale ``superstructure''
to an HMM, which already has a subwavelength ``substructure''. In
such \emph{multiscale HMMs}, schematically shown in Fig.~\ref{FIG:Schematic}d,
VPP propagation is expected to be affected by the superstructure just
as conventional light waves are controlled in photonic crystals. Just
as unprecedented light control can be achieved in PBG structures,
owing to a free choice of geometrical structures (e.g. periodic, coupled-cavity,
quasiperiodic, fractal, etc.) and parameters, unprecedented control
of plasmonic wave propagation should be possible by choosing an appropriate
superstructure geometry in multiscale HMMs. %
\begin{comment}
JS2-old version Since it is known that the extent of light control
in PBG structures is unprecedented, owing to a free choice of geometrical
parameters (e.g. periodic, coupled-cavity, quasiperiodic, fractal,
or other geometries), choosing an appropriate superstructure geometry
in multiscale HMMs should lead to similar control over subwavelength
waves and constitute the design platform for HMM-based elements capable
of bulk plasmonic wave manipulation.%
\end{comment}
{}

In this paper, we demonstrate this multiscale approach by proposing
proof-of-concept designs of Bragg reflectors and Fabry-P{\'e}rot
resonators for high-$k$ bulk plasmons in multilayer metal-dielectric
HMMs. Formation of PBGs in periodic multiscale multilayers is clearly
seen in the numerically calculated Fresnel reflection coefficients
in $k$-space. Breaking the periodicity in the superstructure is shown
to result in cavity resonant modes. Making the structure totally non-periodic
increases the degree of freecom in using PBG effects to control the
dispersion properties of HMMs. In particular, fractal multiscale HMMs
are found to exhibit characteristic self-similar spectral features.
Besides showing that high-$k$ waves can be directly controlled by
PBG effects, the proposed approach is useful in designing HMM-based
devices to engineer and probe the spontaneous emission rate of nearby
atoms in the evanescent-wave domain. 

The paper is organized as follows. In Section \ref{SEC:PROPAGATION},
we review the theoretical background on wave propagation in metal-dielectric
multilayer HMMs, and discuss the dispersion relation of high-$k$
VPP waves in such multilayers. In Section \ref{SEC:MULTISCALE}, we
introduce the concept of multiscale HMMs and show that VPPs can be
manipulated by PBG effects. In particular, we demonstrate Bragg reflection
and Fabry-P{\'e}rot resonances for VPPs in several periodic and nonperiodic
multiscale geometries, including practically realizable designs. Finally,
Section \ref{SEC:CONCLUSIONS} summarizes the paper.

\section{Volume plasmon polaritons in multilayer hyperbolic metamaterials\label{SEC:PROPAGATION}}

Consider a subwavelength periodic metal-dielectric multilayer as shown
in Fig.~\ref{FIG:Schematic}c, where the permittivities of dielectric
and metal are $\epsilon_{d}=\epsilon'_{d}+i\epsilon''_{d}$ ($\epsilon'_{d}>0$)
and $\epsilon_{m}=\epsilon'_{m}+i\epsilon''_{m}$ ($\epsilon'_{m}<0$),
respectively. The layer thicknesses are $d_{d}$ for the dielectric
and $d_{m}$ for the metal, and $\rho\equiv d_{m}/(d_{m}+d_{d})$
denotes the metal filling fraction. 

Any plane wave existing in such a multilayer can have its wave vector
$\mathbf{k}$ represented as a sum of its in-plane component $\bm{\kappa}=k_{x}\hat{\mathbf{x}}+k_{y}\hat{\mathbf{y}}$,
and its out-of-plane component $\mathbf{w}=\pm w\hat{\mathbf{z}}$.
The former is constant across all layers due to the boundary conditions,
so $\kappa=|\bm{\kappa}|$ can be conveniently used as a labeling
parameter for the waves. The out-of-plane component can take the value
$\pm w$ in each layer (denoted by the subscript $m$ or $d$) with
\begin{equation}
w_{m,d}=\sqrt{\left(\frac{\omega}{c}\right)^{2}\epsilon_{m,d}-\kappa^{2}}.\label{eq:w}\end{equation}
Generally we choose the square root of a complex number $\sqrt{z}$
such that $\mathrm{Im}\,\sqrt{z}>0$, taking $\mathrm{Re}\,\sqrt{z}\geq0$
if $\mathrm{Im}\,\sqrt{z}=0$.Neglecting material absorption for now
($\epsilon''_{m,d}=0$), we can see that $w_{d}$ is real for $\kappa<\sqrt{\epsilon_{d}}\omega/c$,
corresponding to propagating waves within the light cone for the dielectric
layers, or purely imaginary otherwise, corresponding to evanescent
waves outside the light cone. For metal layers below the plasma frequency,
$\epsilon{}_{m}<0$, so $w_{m}$ is always imaginary. 

If the layer thicknesses are subwavelength, the effective medium model
is commonly used, and the entire multilayer is regarded as a homogeneous
medium with the permittivity tensor $\hat{\epsilon}_{\eff}=\text{diag}(\epsilon_{x},\epsilon_{y},\epsilon_{z})$,
where\begin{equation}
\epsilon_{x}=\epsilon_{y}=\rho\epsilon_{m}+(1-\rho)\epsilon_{d},\;\epsilon_{z}^{-1}=\rho\epsilon_{m}^{-1}+(1-\rho)\epsilon_{d}^{-1}.\label{eq:epsxz}\end{equation}

In such an extremely anisotropic medium, the expression for $w$ of
a $p$-polarized wave (for which surface plasmons can propagate along
metal-dielectric interfaces) is \cite{0Saarinen}\begin{equation}
w_{\eff}=\sqrt{\left(\frac{\omega}{c}\right)^{2}\epsilon_{x}-\frac{\epsilon_{x}}{\epsilon_{z}}\kappa^{2}}.\label{eq:wHMM}\end{equation}
For layered HMMs it is typical that $\epsilon_{x}<0$ and $\epsilon_{z}>0$
(again, neglecting absorption for the moment). Then, we see that the
second term under the square root in Eq.~\eqref{eq:wHMM} becomes
negative and overrules the first term for large enough $\kappa$.
So the entire expression under the root becomes positive. Thus, the
waves change from evanescent (imaginary $w$) to propagating (real
$w$) at $\kappa=\kappa_{c}$ defined as $w_{\eff}(\kappa_{c})=0$. 

Continuing to neglect absorption, we consider the expression of the
Fresnel reflection coefficient for a boundary between a homogeneous
dielectric and a medium described by the permittivity tensor in Eq.~\eqref{eq:epsxz}
\cite{ourFocusOE13}, \begin{equation}
R_{\eff}=\frac{w_{d}\epsilon_{x}-w_{\eff}\epsilon_{d}}{w_{d}\epsilon_{x}+w_{\eff}\epsilon_{d}}.\label{eq:rHMM}\end{equation}
In the region of large real $\kappa$, for which $w_{d}$ is imaginary,
we see that real (rather than imaginary) $w_{\eff}$ causes $R_{\eff}$
to acquire a non-vanishing imaginary part \cite{0ourHMMPRA}. In other
words, $\mathrm{Im}\, R_{\eff}(\kappa)$ is non-zero for those values
of $\kappa$ that correspond to propagating waves in the effective
medium. 

\begin{figure}[tb]
\begin{centering}
\includegraphics[bb=0bp 0bp 310bp 55bp,clip,width=1\columnwidth]{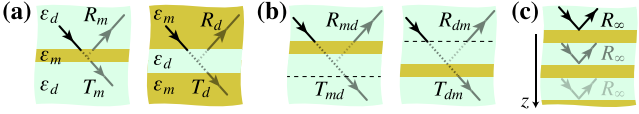}
\par\end{centering}

\caption{(Color online) Illustraton of the Fresnel reflection and transmission
coefficient derivation: (a) for a single metal or dielectric layers
as in Eq.~\eqref{eq:Airy}; (b) for a metal-dielectric bilayer as
in Eq.~\eqref{eq:Airy_bilayer}; (c) for a semi-infinite periodic
metal-dielectric multilayer as in Eq.~\eqref{eq:Airy_inf}. \label{FIG:bilayers}}
\end{figure}

\begin{figure}[tb]
\begin{centering}
\includegraphics[width=1\columnwidth]{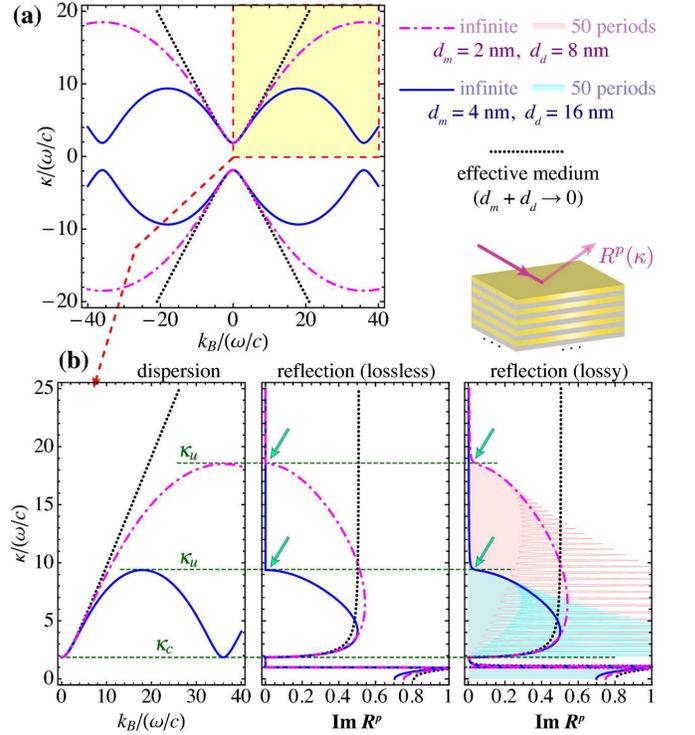}
\par\end{centering}

\caption{(Color online) (a) Dispersion relation for high-$k$ Bloch waves in
an infinite metal-dielectric HMM comprising alternating Ag/epoxy layers
($\epsilon_{m}=-30.1+i\epsilon_{m}''$, $\epsilon_{d}=2.72$ for $\lambda=720$
nm \cite{0LoslessHMM}) with $\rho=0.2$ and different layer thicknesses.
(b) An enlarged view of a part of that dispersion relation along with
the imaginary part of the Fresnel reflection coefficient, $\mathrm{Im\,}R_{\infty}^{p}$,
for a semi-infinite effective multilayer (see inset) without losses
($\epsilon_{m}''=0$) and with losses ($\epsilon_{m}''=0.41$) \cite{0ourHMMPRA}.
The dotted lines denote the limiting case of the homogeneous effective
anisotropic medium {[}Eq.~\eqref{eq:epsxz}{]}, corresponding to
a multilayer with infinitely thin layers. The horizontal dashed lines
denote the VPP band edges, and faint green arrows highlight the differences
in $\mathrm{Im\,}R_{\infty}^{p}$ for lossless vs.~real metal. The
shaded areas in the rightmost plot correspond to a finite (50-period)
multilayer as opposed to an infinite one. \label{FIG:character}}
\end{figure}

This correspondence is physically significant, and can be extended
to the case of actual multilayers. On the one hand, the dispersion
relation of propagating Bloch waves in an infinite periodic metal-dielectric
multilayer can be determined from the transfer matrix method \cite{BookYarivYeh}.
The transfer matrix of one period can be written as \begin{equation}
M_{1}=\frac{1}{T_{m}}\left[\begin{array}{cc}
T_{m}^{2}-R_{m}^{2} & R_{m}\\
-R_{m} & 1\end{array}\right]\left[\begin{array}{cc}
e^{iw_{d}d_{d}} & 0\\
0 & e^{-iw_{d}d_{d}}\end{array}\right],\label{eq:onematrix}\end{equation}
where the reflection and transmission coefficients of a metal layer
(see Fig.~\ref{FIG:bilayers}a) are given by the Airy formulas, \begin{equation}
R_{m}=r_{dm}+\frac{t_{dm}r_{md}t_{md}e^{2iw_{m}d_{m}}}{1-r_{md}^{2}e^{2iw_{m}d_{m}}},\;\; T_{m}=\frac{t_{dm}t_{md}e^{iw_{m}d_{m}}}{1-r_{md}^{2}e^{2iw_{m}d_{m}}};\label{eq:Airy}\end{equation}
the coefficients $r_{dm,md}$ and $t_{dm,md}$ are the interface coefficients
for $p$-polarized waves determined by the Fresnel formulas, \begin{equation}
\begin{gathered}r_{md}=\frac{w_{m}\epsilon_{d}-w_{d}\epsilon_{m}}{w_{m}\epsilon_{d}+w_{d}\epsilon_{m}},\; r_{dm}=\frac{w_{d}\epsilon_{m}-w_{m}\epsilon_{d}}{w_{d}\epsilon_{m}+w_{m}\epsilon_{d}};\\
t_{md}=\frac{2w_{m}\sqrt{\epsilon_{m}\epsilon_{d}}}{w_{m}\epsilon_{d}+w_{d}\epsilon_{m}},\; t_{dm}=\frac{2w_{d}\sqrt{\epsilon_{d}\epsilon_{m}}}{w_{d}\epsilon_{m}+w_{m}\epsilon_{d}}.\end{gathered}
\label{eq:Fresnel}\end{equation}
and $w_{m,d}$ are given by Eq.~\eqref{eq:w}. According to Bloch's
theorem, the normal wave vector component $k_{B}$ of the propagating
Bloch wave with tangential wave vector component $\kappa$ ($k^{2}=k_{B}^{2}+\kappa^{2}$)
is determined as $\cos[k_{B}(d_{m}+d_{d})]=(\mathrm{Tr}\, M_{1})/2$,
resulting in a well-known dispersion relation \cite{blFengOE05,blOrenOE11,0LoslessHMM}\begin{equation}
\begin{gathered}\cos\left[k_{B}(d_{m}+d_{d})\right]=\cos(w_{m}d_{m})\cos(w_{d}d_{d})\\
-\frac{1}{2}\left(\frac{\epsilon_{m}w_{d}}{\epsilon_{d}w_{m}}+\frac{\epsilon_{d}w_{m}}{\epsilon_{m}w_{d}}\right)\sin(w_{m}d_{m})\sin(w_{d}d_{d}).\end{gathered}
\label{eq:Bloch}\end{equation}
The solution of this equation in the wave vector space defines a band
of propagating high-$k$ VPP waves (see Fig.~\ref{FIG:character}a),
which exists within certain limits $\kappa_{c}<\kappa<\kappa_{u}$.
The lower band edge $\kappa_{c}$ is determined by the above mentioned
condition $w_{\eff}(\kappa_{c})=0$, and coincides with the prediction
of the effective medium theory. Conversely, the upper band edge $\kappa_{u}\propto(d_{m}+d_{d})^{-1}$
is associated with the breakdown of that approximation due to the
finite layer thickness \cite{0ourHMMPRA}. 

On the other hand, the reflection coefficient of a semi-infinite multilayer
HMM can also be analytically determined. Applying the Airy formulas
\eqref{eq:Airy} one more time to a stack of two layers (metal and
dielectric, see Fig.~\ref{FIG:bilayers}b) lets us express the transmission
and reflection of such bilayers in terms of the single-layer reflection
and transmission coefficients as \cite{ourPRA10} \begin{equation}
R_{md}=R_{m},\enskip R_{dm}=R_{m}e^{2iw_{d}d_{d}},\enskip T_{md}=T_{dm}=T_{m}e^{iw_{d}d_{d}},\label{eq:Airy_bilayer}\end{equation}
with $R_{m}$ and $T_{m}$ given by Eq.~\eqref{eq:Airy}. Since a
semi-infinite structure remains unchanged when its outermost period
is removed, its reflection coefficient $R_{\infty}$ must satisfy
a quadratic equation\begin{equation}
R_{\infty}=R_{md}+\frac{T_{md}R_{\infty}T_{dm}}{1-R_{\infty}R_{dm}}.\label{eq:Airy_inf}\end{equation}
Solving this equation and choosing the solution for which the wave
would be decaying, rather than growing, as $z$ increases (see Fig.~\ref{FIG:bilayers}c)
we can determine $R_{\infty}$ \cite{ourOC13}. 

In the absence of losses and for $\kappa>(\omega/c)\sqrt{\epsilon_{d}}$,
we can see that $w_{m,d}$ are purely imaginary, and consequently,
$R_{md,dm}$ and $T_{md,dm}$ are purely real {[}see Eqs.~\eqref{eq:Airy}--\eqref{eq:Fresnel}{]}.
Then it can be shown that the existence condition for VPPs in the
form $|\mathrm{Tr}\, M_{1}|\leq2$ coincides with the condition $\mathcal{D}\leq0$,
where $\mathcal{D}$ is the discriminant of the quadratic equation
\eqref{eq:Airy_inf}, ensuring that its roots become complex even
though its coefficients are real. This generalizes the relation between
non-zero imaginary part of the reflection coefficient at an interface
and the existence of propagating waves beyond the interface in the
evanescent-wave domain from the case of a single interface (as seen
in Eqs.~\eqref{eq:wHMM}--\eqref{eq:rHMM} above and as elaborated
in Appendix~\ref{sec:APPENDIX}) to the case of a semi-infinite periodic
multilayer HMM. Namely, we find that the imaginary part of $R_{\infty}$
is non-zero within the VPP band found by Bloch's theorem. Indeed,
Fig.~\ref{FIG:character}b shows that the range where $\mathrm{Im\,}R_{\infty}\neq0$
is exactly $\kappa_{c}<\kappa<\kappa_{u}$, where propagating high-$k$
VPP waves were shown to exist. Elsewhere in the high-$\kappa$ range,
$\mathrm{Im\,}R_{\infty}=0$ and no propagating solutions are allowed. 

Even though rigorously derived for lossless, semi-infinite multilayers,
this criterion is still a useful one if losses are present ($\epsilon''_{m,d}\ne0$).
Figure \ref{FIG:character}b shows that the abrupt appearance of non-zero
$\mathrm{Im\,}R_{\infty}$ at the band edges is smeared \cite{0ourHMMPRA}
because the sharp distinction between evanescent and propagating waves
can no longer be made if the materials are lossy. However, the general
argument still persists that $\mathrm{Im\,}R_{\infty}$ is significantly
non-zero when the Bloch solutions of the dispersion equation are propagating. 

Moreover, including losses makes it possible to generalize the relation
between non-zero $\mathrm{Im\,}R_{\infty}$ and the existence of propagating
waves inside a finite multilayer structure. Note first that for lossless
\emph{finite} multilayer structures with $\kappa>(\omega/c)\sqrt{\epsilon_{d}}$
we see that all quantities entering the transfer matrix $M_{1}$ are
real. Hence, the reflection coefficient of such a multilayer structure
must be real, too, with the exception of a discrete set of poles where
the reflection coefficient diverges and its phase becomes indeterminate;
these poles are known to signify the presence of guided modes inside
the multilayer. As the number of layers in the structure increases,
the number of poles grows accordingly, but it is only in the limit
of an \emph{infinite} multilayer that the transition from a discrete
set to a continuous band can occur, as shown above. When it occurs,
modes that are guided along the layers in a finite multilayer acquire
a real $z$-component of the wave vector {[}$k_{B}$ as given by Eq.~\eqref{eq:Bloch}{]}
and become propagating {}``through the bulk'' of an infinite multilayer;
it is for this reason that we refer to these waves as \emph{volume}
plasmon polaritons. 

Thus, in the truly lossless case, finite multilayers only support
surface waves with a discrete set of $\kappa>(\omega/c)\sqrt{\epsilon_{d}}$,
whereas infinite multilayers can additionally support bulk propagating
waves (VPPs) in a continuous range of $\kappa$. However, the presence
of losses (even very minor ones) regularizes this opposition, transforming
each discrete point into a narrow peak where $\mathrm{Im\,}R(\kappa)\neq0$.
When there are many layers in the structure, some of these peaks typically
merge into a continuous band (see Fig.~\ref{FIG:character}b), which
is seen to approach the dependence $\mathrm{Im\,}R_{\infty}(\kappa)$
as the number of layers increases. Within this VPP band, there are
waves inside the multilayer HMM that are quasi-propagating in the
sense that (i) their propagating character is primarily determined
by the infinite-structure dispersion relation and is only weakly influenced
by the number of layers in the structure, (ii) they undergo a much
weaker attenuation than they would undergo in any homogeneous isotropic
medium, and (iii) they become less attenuated if losses are lowered.
In contrast, waves outside the VPP band remain strongly evanescent
regardless of whether material losses are present.

Hence we will be using the existence of $\mathrm{Im}[R(\kappa)]$
as a {}``footprint,'' providing evidence for the existence of high-$k$
waves, or VPPs, in a range of HMM multilayer structures.

\section{Multiscale hyperbolic metamaterials\label{SEC:MULTISCALE}}

\begin{figure}[tb]
\begin{centering}
\includegraphics[width=1\columnwidth]{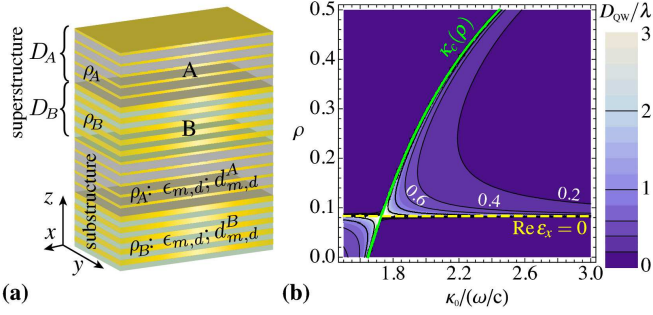}
\par\end{centering}

\caption{(Color online) (a) Schematics of a multiscale HMM with periodic geometry,
showing the division into superstructure and substructure. (b) Dependence
of the QW layer thickness $D_{\text{QW}}/\lambda$ on filling fraction
$\rho$ and the target PBG location $\kappa_{0}$ according to Eq.~\eqref{eq:super}.
The solid green line denotes the lower VPP band edge $\kappa_{c}(\rho)$,
and the dashed yellow line marks the boundary of the HMM regime ($\mathrm{Re}[\epsilon_{x}]=0$).
\label{FIG:quarterwave}}
\end{figure}

Since VPPs are bulk Bloch plasmonic waves with propagating character,
they have to be subject to the PBG effects just as any other kind
of propagating waves. A straightforward idea is to apply these PBG
effects to modulate the properties of a subwavelength multilayer HMM
(in particular, the metal filling fraction $\rho$) on a larger length
scale (Fig.~\ref{FIG:Schematic}d). To distinguish the two scales,
we will refer to the coarser, wavelength-scale modulation as the {}``superstructure''
consisting of several \emph{superlayers}; Fig.~\ref{FIG:quarterwave}a
displays two kinds of such superlayers (denoted A and B) with thicknesses
$D_{A,B}$ and filling fraction $\rho_{A,B}$, respectively. The finer
subwavelength periodic metal-dielectric structure within each superlayer,
which gives rise to HMM properties, is called the {}``substructure''.
Thus, each superlayer contains a certain number of subperiods, $N_{A,B}=D_{A,B}/(d_{m}+d_{d})$,
or twice as many \emph{sublayers}. 

The resulting multiscale multilayer is expected to exert the same
degree of control over VPP waves as the corresponding photonic multilayers
control propagation of conventional electromagnetic waves. Hence,
a periodic arrangement of superlayers, where the layers denoted by
A and B simply alternate, should result in a band gap for VPPs. In
a further analogy with photonic multilayers, this band gap should
forbid the propagation of waves with values of $\kappa$ around a
certain mid-gap value $\kappa_{0}$, for which the superlayers are
close to be quarter-wave (QW), i.e., their optical thickness should
be close to one quarter of the wavelength that corresponds to $w_{\eff}(\kappa_{0})$.
In the remainder of this section, we will investigate the influence
of the superstructure geometry on the VPP propagation in a variety
of multiscale HMMs.

\subsection{Bragg reflectors and Fabry-P{\'e}rot resonators \protect \\
with thick-layer superstructure}

In order to demonstrate the multiscale concept, we will first consider
structures where the superstructure and substructure length scales
are clearly separated, i.e., $\lambda/4\simeq D_{A,B}\gg d_{m,d}$
(or $N_{A,B}\gg1$). Keeping in mind that the wavelength of high-$k$
VPPs, $\lambda\simeq2\pi/k$, can be anomalously small compared to
the vacuum wavelength $\lambda_{0}=\omega/c$, it can be expected
that the target $\kappa_{0}$ should be no more than several times
larger than $\omega/c$. To overcome this apparent contradiction,
we determine the thickness $D_{\text{QW}}$ of a {}``model'' homogeneous
QW layer made of the effective medium {[}see Eq.~\eqref{eq:epsxz}{]}
for a given $\kappa_{0}$. From Eq.~\eqref{eq:wHMM}, \begin{equation}
D_{\text{QW}}=\frac{\lambda}{4}\left(\mathrm{Re}\left[\sqrt{\frac{\omega^{2}}{c^{2}}\epsilon_{x}-\frac{\epsilon_{x}}{\epsilon_{z}}\kappa_{0}^{2}}\right]\right)^{-1}\label{eq:super}\end{equation}
The dependence of $D_{\text{QW}}$ on $\rho$ and $\kappa_{0}$ is
shown in Fig.~\ref{FIG:quarterwave}b. It is seen that to be able
to form a PBG well into the high-$\kappa$ range and still have a
clear separation of length scales ($D_{\text{QW}}\gg d_{m,d}$), one
should use filling fractions slightly above 0.08, for which $\epsilon_{x}\lesssim0$.
Otherwise, we see that $D_{\text{QW}}\ll\lambda$ unless $\kappa_{0}$
is very close to the lower VPP band edge ($\kappa\gtrsim\kappa_{c}$).

\begin{figure*}
\begin{centering}
\includegraphics[width=1\textwidth]{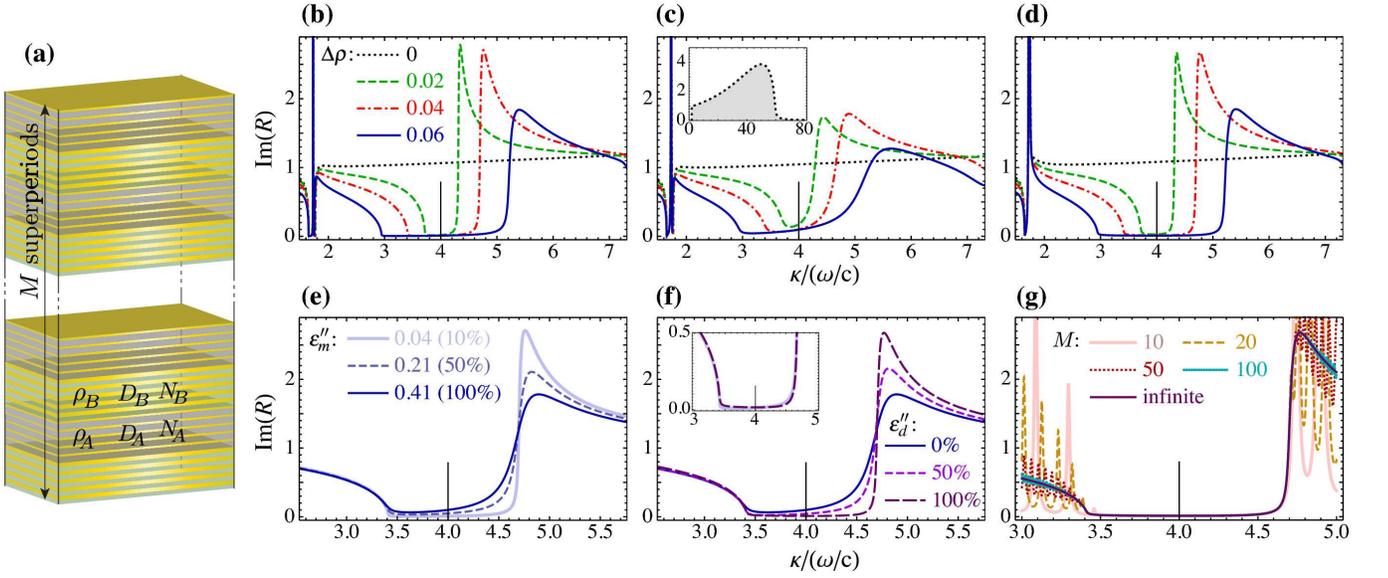}
\par\end{centering}

\caption{(Color online) Characterization of a multiscale Bragg reflector for
high-$k$ waves shown schematically in (a) with the default geometry
comprising a superstructure with $M=10^{4}$ periods of HMM layers
with $(\rho_{A}+\rho_{B})/2=0.12$, the substructure period $d_{m}+d_{d}=4$~nm,
and superlayer thickness $N_{A,B}$ determined from Eq.~\eqref{eq:super}.
\emph{Top row}: the function $\mathrm{Im}[R(\kappa)]$ for a varying
depth of superstructure modulation ($\Delta\rho=\rho_{A}-\rho_{B}$)
for (b) 10\% losses in silver ($\epsilon''_{m}=0.04$), (c) full losses
in silver ($\epsilon''_{m}=0.41$), and (d) full losses in silver
compensated by gain in the epoxy layers as described in \cite{0LoslessHMM}
($\epsilon''_{m}=0.41$ and $\epsilon''_{d}=-0.04$); the dotted line
shows the structure with no superstructure ($\rho_{A}=\rho_{B}=0.12$),
with the full high-$\kappa$ band shown in the inset of (c). \emph{Bottom
row}: the dependence of $\mathrm{Im}[R(\kappa)]$ for the structure
with $\rho_{A}=0.1$, $\rho_{B}=0.14$ and $N_{A}=29$, $N_{B}=16$
near the gap region (e) for varying amount of loss in silver $\epsilon''_{m}$
for $\epsilon''_{d}=0$, (f) for varying amount of gain in dielectric
$\epsilon''_{d}$ for $\epsilon''_{m}=0.41$, and (g) for a varying
number of superperiods $M$. The inset in (f) compares the cases of
10\% losses without gain and 100\% losses with gain. \label{FIG:mirror}}
\end{figure*}

Therefore, the design of a Bragg reflector for VPPs involves choosing
two values of $\rho$ for the superlayers ($\rho_{A}$ and $\rho_{B}$),
and then using Eq.~\eqref{eq:super} to determine the superlayer
thicknesses $D_{A,B}$. Following the example of Ni \emph{et al.}\ \cite{0LoslessHMM},
and choosing silver and epoxy as metal and dielectric materials, respectively
($\epsilon_{m}=-30.1+0.41i$, $\epsilon_{d}=2.72$ for $\lambda=720$
nm), we choose $\rho_{A}=0.1$ and $\rho_{B}=0.14$. To form a band
gap at a target $\kappa_{0}=4\omega/c$, we arrive at $D_{A}=116.4$
nm and $D_{B}=65.8$ nm, approximately corresponding to $N_{A}=29$
and $N_{B}=16$ metal-dielectric bilayers with $d_{m}+d_{d}=4$ nm
thickness (Fig.~\ref{FIG:mirror}a).

Such layers would be very difficult to fabricate, and are so thin
that it is unlikely bulk optical constants could be used to characterize
them. Nonetheless, as an initial analysis we theoretically characterize
these nominal structures (Fig.~\ref{FIG:mirror}b--g) to help identify
some of the essential physics. Considering first the semi-infinite
superstructure ($M=10^{4}$ periods in practical calculations) to
suppress pass-band states, and artificially lowering the imaginary
part of the metal permittivity to 10\% of its actual value, we clearly
see a range of very low $\mathrm{Im}\, R$ around the target midgap
$\kappa_{0}=4\omega/c$ (Fig.~\ref{FIG:mirror}b), signifying inhibited
VPP propagation as described in Section~\ref{SEC:PROPAGATION} above.
This range, or stop band, is seen to widen as the modulation depth
of the filling fraction $\rho$ increases, which is characteristic
of a PBG opening and confirms that VPPs indeed undergo Bragg reflection
in a periodic-superstructure multiscale HMM.

Restoring the amount of losses in metal to their actual value (Fig.~\ref{FIG:mirror}c),
we see that the band gap is less pronounced but nevertheless quite
visible. Compensating the loss in silver by incorporating optical
gain in the dielectric layers of the HMM (putting $\epsilon_{d}=2.72-0.04i$,
as was recently envisaged by Ni \emph{et al.}\ \cite{0LoslessHMM})
restores the pronounced character of the band gap (see Fig.~\ref{FIG:mirror}d).
The effects of loss and gain on the band gap visibility are additionally
highlighted in Fig.~\ref{FIG:mirror}e--f where it is shown that
adding gain to the dielectric layers indeed results in partial loss
compensation. Since the working filling fractions are around 0.1,
so that $d_{d}\approx10d_{m}$, the negative imaginary component of
$\epsilon_{d}$ in the range of 0.04 is sufficient to compensate the
loss in silver corresponding to $\epsilon''_{m}=0.41$. Indeed, the
inset in Fig.~\ref{FIG:mirror}f demonstrates that the gap profile
for the gain-compensated HMM is almost identical with that for the
structure where loss is artificially reduced to 10\%. Finally, Fig.~\ref{FIG:mirror}g
shows that reducing the number of superperiods does not change the
location of the stop band for VPPs but strongly modifies the propagation
of such waves in the surrounding {}``pass band'' of the high-$\kappa$
range, featuring multiple band edge-like resonances.

\begin{figure}[tb]
\begin{centering}
\includegraphics[width=1\columnwidth]{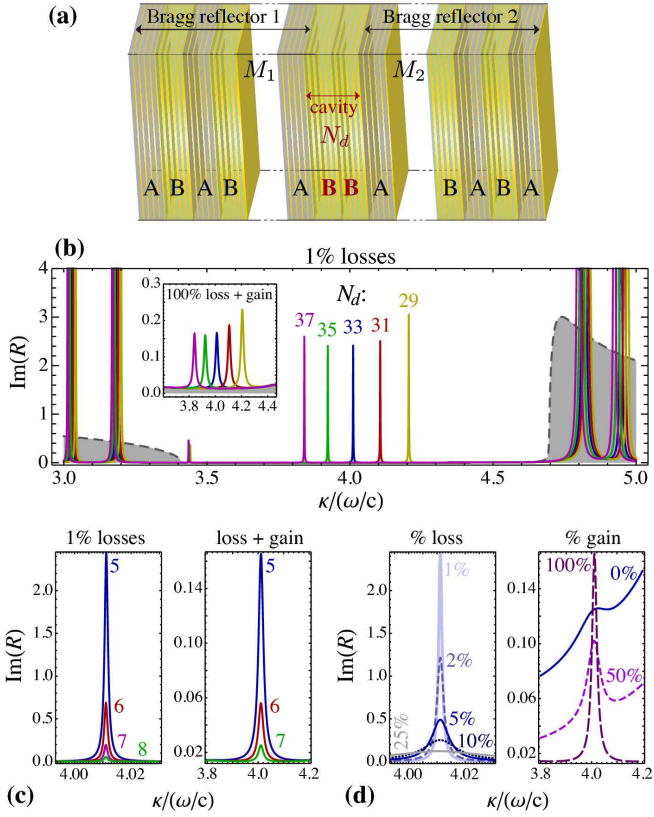}
\par\end{centering}

\caption{(Color online) Characterization of a multiscale Fabry-P\'{e}rot resonator
for VPPs. (a) Schematics of the structure. (b) Dependence $\mathrm{Im}[R(\kappa)]$
for different thickness of the defect layer expressed in the number
of superperiods $N_{d}$ for a superstructure with the same parameters
as in Fig.~\ref{FIG:mirror} but with $M_{1}=M_{2}=5$ and 1\% losses
($\epsilon''_{m}=0.004$); the inset shows the excerpt of the same
dependence for full losses in metal compensated by gain in the dielectric
($\epsilon''_{m}=0.41$ and $\epsilon''_{d}=-0.04$). Also shown are
the enlarged views of the cavity resonance peak under varying conditions
in structures with (c) varying number of superperiods $M_{1}=M_{2}$
and (d) varying degree of loss and gain in the constituent materials.\label{FIG:resonator}}
\end{figure}

Stacking a periodic Bragg reflector with its mirror image forms a
structure with geometry $\mathrm{ABAB}\ldots\mathrm{A\mathbf{BB}A}\ldots\mathrm{ABAB}$
(Fig.~\ref{FIG:resonator}a), creating a half-wave defect or cavity
(``$\mathbf{BB}$'') in an otherwise periodic structure. By analogy
with photonic multilayers, such a structure should function as a Fabry-P{\'e}rot
resonator for VPPs, creating a narrow band of $\kappa\approx\kappa_{\text{res}}$
where the plasmon energy gets trapped in the cavity and the waves
can tunnel through the structure despite the presence of a band gap.
This feature should be observed as a sharp peak of $\mathrm{Im}[R(\kappa_{\text{res}})]$
inside the forbidden gap. Varying the thickness of the cavity layer
(by adjusting the number of subperiods in it), it should be possible
to tune the location of $\kappa_{\text{res}}$ across the band gap. 

Such behaviour is indeed observed in Fig.~\ref{FIG:resonator}. The
peak location is seen to shift as additional substructure periods
are removed from or added to the defect layer, varying its thickness
(Fig.~\ref{FIG:resonator}b). Note the stark contrast between the
marked shift of the central resonance peak and nearly absent shift
of the remaining resonances in the wavevector spectra. This means
that the central peak is a Fabry-P{\'e}rot resonance whereas the
remaining peaks are not related to the cavity layer%
\begin{comment}
 JS9-deleted: and are most likely of the band edge nature%
\end{comment}
{}. For VPPs corresponding to the Fabry-P{\'e}rot resonance, it can
be concluded that they are localized in the defect and guided within
the $x$-$y$ plane. 

Note that a much smaller number of periods in the superstructure is
necessary for the appearace of the resonance peak than for the appearance
of the band gap (see Fig.~\ref{FIG:mirror}). For a larger number
of periods in the superstructure it is seen that the guided high-$k$
VPP waves decouple from the incident wave, making them harder to characterize
or interact with (Fig.~\ref{FIG:resonator}c). Also, similar to what
we could see in Fig.~\ref{FIG:mirror}, absorption in metal is highly
detrimental: without compensation, the resonance peaks all but vanish
when the imaginary part of $\epsilon_{m}$ reaches 10\% of its actual
value. However, in presence of gain the peaks are seen to reappear
even with full metal losses (Fig.~\ref{FIG:resonator}d); the peaks
are notably broadened but their location is not affected (see the
inset in Fig.~\ref{FIG:resonator}b).

\subsection{Bragg reflectors and Fabry-P{\'e}rot resonators \protect \\
with thin-layer superstructure}

The approach of the previous subsection has a didactic advantage,
with its easy separation of superstructure and substructure length
scales; this makes the multiscale features that arise easy to understand.
However, structures with $d_{m}+d_{d}=4$ nm, and indeed with $d_{m}\simeq0.5$
nm, cannot be easily fabricated and, even if they could, the optical
response of such thin layers would not likely be described by bulk
dielectric constants. Further, even were fabrication possible it was
shown that absorption is very detrimental to PBG effects, requiring
either low-loss plasmonic materials \cite{BetterPlasmonics} or loss
compensation means \cite{0LoslessHMM} for the effects to be observed.
These disadvantages are worsened by the need for structures consisting
of hundreds (if not thousands) of sublayers, leading even the most
optimistic to despair of fabrication at any time in the near future.
As a result, thick-superlayer structures for VPPs in multilayer HMMs
can only be considered as proof-of-concept structures, and are not
viable from a practical point of view.

\begin{figure*}
\begin{centering}
\includegraphics[width=1\textwidth]{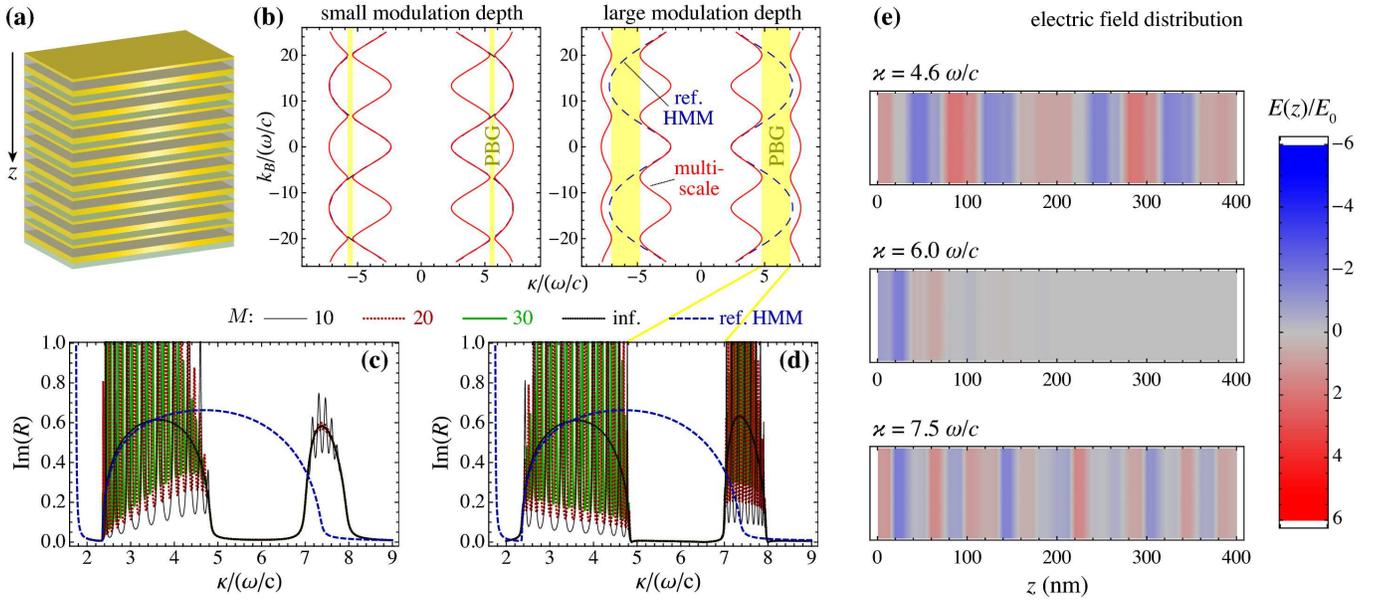}
\par\end{centering}

\caption{(Color online) Characterization of a Bragg reflector for high-$k$
waves comprising a superstructure of 10 to 30 layers, each consisting
of one subperiod ($N_{A}=N_{B}=1$) with $\rho_{A}=0.60$, $\rho_{B}=0.34$,
and $d_{m}+d_{d}=20$ nm. (a) Schematic of the structure; (b) Dispersion
relation similar to Fig.~\ref{FIG:character} for an infinite double-periodic
structure for small and large filling fraction modulation $\Delta\rho$.
(c--d) $\mathrm{Im}\, R(\kappa)$ for full metal losses (c) without
and (d) with gain in the dielectric. The dashed line shows the reference
VPP band from a HMM with average $\rho_{A}=\rho_{B}=0.473.$\textbf{
}(e) Electric field distribution in the structure with 20 periods
for $\kappa c/\omega$\textbf{ }equal to 4.6 (below the band gap);
6.0 (inside the band gap); 7.5\textbf{ }(above the band gap).\textbf{\label{FIG:mirror-realizable}}}
\end{figure*}

To alleviate these fabricational challenges, we consider here the
other extreme of the multiscale concept and analyze the structures
where superlayers consist of just one subperiod ($N_{A}=N_{B}=1$),
allowing the metal and dielectric sublayers to be only moderately
thin. In this regime, the subwavelength approximations will certainly
fail \cite{0ourHMMPRA}, so Eq.~\eqref{eq:super} can no longer be
regarded as an accurate prediction of a QW layer thickness. Instead,
the structure can be considered as a periodic multilayer with a more
complicated unit cell containing four layers (a double-periodic structure,
see Fig.~\ref{FIG:mirror-realizable}a). Calculating its transfer
matrix in a similar manner to Eq.~\eqref{eq:onematrix} and applying
Bloch's theorem along the lines of Section \ref{SEC:PROPAGATION},
one can obtain the high-$k$ dispersion relation of such a double-periodic
structure with infinite number of periods. One can see (Fig.~\ref{FIG:mirror-realizable}b)
that if the filling fraction difference $\Delta\rho=\rho_{A}-\rho_{B}$
is non-zero, the VPP band splits into two, leaving a gap between them,
which widens as $\Delta\rho$ increases. Since we are no longer restricted
by the condition $D_{A,B}\gg d_{m.d}$, we can consider a structure
with thicker sublayers (e.g., $d_{m}+d_{d}=20$ nm), which would be
far more realistic for fabrication. We are also free to work with
a greater contrast of $\rho$, and, as seen in Fig.~\ref{FIG:mirror-realizable}b,
a prominently wide band gap can be achieved for an example structure
with $\rho_{A}=0.60$ and $\rho_{B}=0.34$. The gap location is now
in the higher-$\kappa$ part of the wavevector space, around $\kappa\simeq5\ldots7\omega/c$,
and it turns out that Eq.~\eqref{eq:super} is still able to give
a meaningful estimate of the gap location, predicting the mid-gap
$\kappa$ to be $6.75\omega/c$. 

Figure \ref{FIG:mirror-realizable}c--d shows the characterization
of such a multiscale HMM with different number of superperiods. We
can see that a PBG for VPPs does form at the predicted location with
clear separation between pass bands and stop bands with as few as
several tens of superlayers. Together with practically achievable
values for the sublayer thicknesses, this makes the whole structure
much more promising for experimental realization than the thick-superlayer
counterparts. Finally, it can be seen that the PBG is still clearly
pronounced with the realistic account for the losses in silver (Fig.~\ref{FIG:mirror-realizable}c).
To make sure that the feature seen in Fig.~\ref{FIG:mirror-realizable}c
is actually a PBG for VPPs seen in Fig.~\ref{FIG:mirror-realizable}b,
we compare the electric field distribution at three different $\kappa$,
namely, below, inside, and above the band gap (Fig.~\ref{FIG:mirror-realizable}e).
The field distribution shows the evanescent vs.\ extended character
of the waves inside the HMM for $\kappa$ inside vs.\ outside of
the band gap, respectively.

\begin{figure*}
\begin{centering}
\includegraphics[width=1\textwidth]{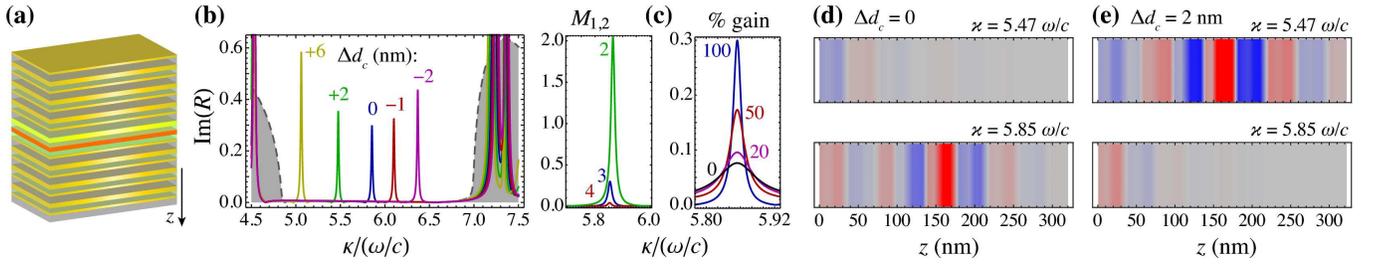}
\par\end{centering}

\caption{(Color online) Characterization of a Fabry-P\'{e}rot resonator for
high-$k$ waves comprising a superstructure with $M_{1}=M_{2}=3$
periods of two superlayers with same substructure as in Fig.~\ref{FIG:mirror-realizable}.
(a) Schematic of the structure, highlighting the cavity region and
variable-thickness metal layer in it; (b) $\mathrm{Im}\, R(\kappa)$
in presence of gain for varying thickness of the central metal layer,
$\Delta d_{c}$ being the thickness adjustment compared to the ideal
half-wave defect; (c) $\mathrm{Im}\, R(\kappa)$ for 100\% of metal
losses and gain compensation in the dielectric. (c) Enlarged view
of the peak for varying number of superperiods and degree of gain.
The pictures at the bottom show the field distribution in the structures
with (d) \textbf{$\Delta d_{c}=0$} and (e) \textbf{$\Delta d_{c}=2$}
nm: top plots, $\kappa=5.47\omega/c$ (on-peak for \textbf{$\Delta d_{c}=2$}
nm); bottom plots, $\kappa=5.85\omega/c$ (on-peak for \textbf{$\Delta d_{c}=0$}).
The color scale is the same as in Fig.~\ref{FIG:mirror-realizable}e.
\label{FIG:resonator-realizable}}
\end{figure*}

By adding gain to the dielectric layers, the gap becomes even more
clearly defined, approaching abrupt band edges characteristic for
the lossless case (see Fig.~\ref{FIG:character}b). This near-total
loss compensation makes this structure a practically realizable candidate
for a Fabry-P{\'e}rot resonator for VPPs. As before, a basic half-wave
defect is formed by repeating one of superlayers twice, or simpler,
by stacking a 3-superperiod structure (ABABAB) with its mirror image
(BABABA), as seen in Fig.~\ref{FIG:resonator-realizable}a. Rather
than varying the number of sublayers in the cavity region (which is
not possible since $N_{A}=N_{B}=1$ and $N_{d}=2$), the resonance
location can be tuned by adjusting the thickness of one of the metal
layers in the cavity region. Indeed, the characterization in Fig.~\ref{FIG:resonator-realizable}b
shows the possibility of moving the VPP resonance peak across the
band gap. The field distribution calculations show that when $\kappa$
matches the peak in the dependence $\mathrm{Im}\, R(\kappa)$, the
structure features a Fabry-P{\'e}rot resonant mode localized near
the cavity layer, whereas elsewhere in the band gap the waves in the
HMM remain evanescent.

Figure \ref{FIG:resonator-realizable}c shows the dependence of the
peak profile on the number of superperiods in the Bragg mirrors surrounding
the cavity. As before, increasing this number beyond 4 makes the peak
vanish by decoupling the localized VPP mode from the incident wave,
and reducing the amount of gain leads to broadening of the peak, again
to the point of vanishing, due to the decrease in its \emph{Q}-factor. 

Overall, we see that thin-superlayer multiscale HMMs can exert the
same PBG behavior on VPPs as their thick-superlayer counterparts,
but with a much smaller number of moderately thin layers, making such
structures much more feasible for experimental realization. As a price
to pay, the gap and/or resonance position can no longer be smoothly
tuned by varying the number of subperiods in the superstructure; other
methods such as departing from the QW condition in selected layers
have to be used instead. We also see that for this thin-superlayer
structure the gain compensation is quite successful in increasing
the visibility of PBG effects. However, we refrain from far-reaching
conclusions based on this result because the behavior of VPPs in presence
of gain strong enough to fully compensate (or even overcompensate)
losses needs further investigation.

\subsection{Multi-gap and fractal Cantor multiscale HMMs}

\begin{figure}[b]
\begin{centering}
\includegraphics[width=1\columnwidth]{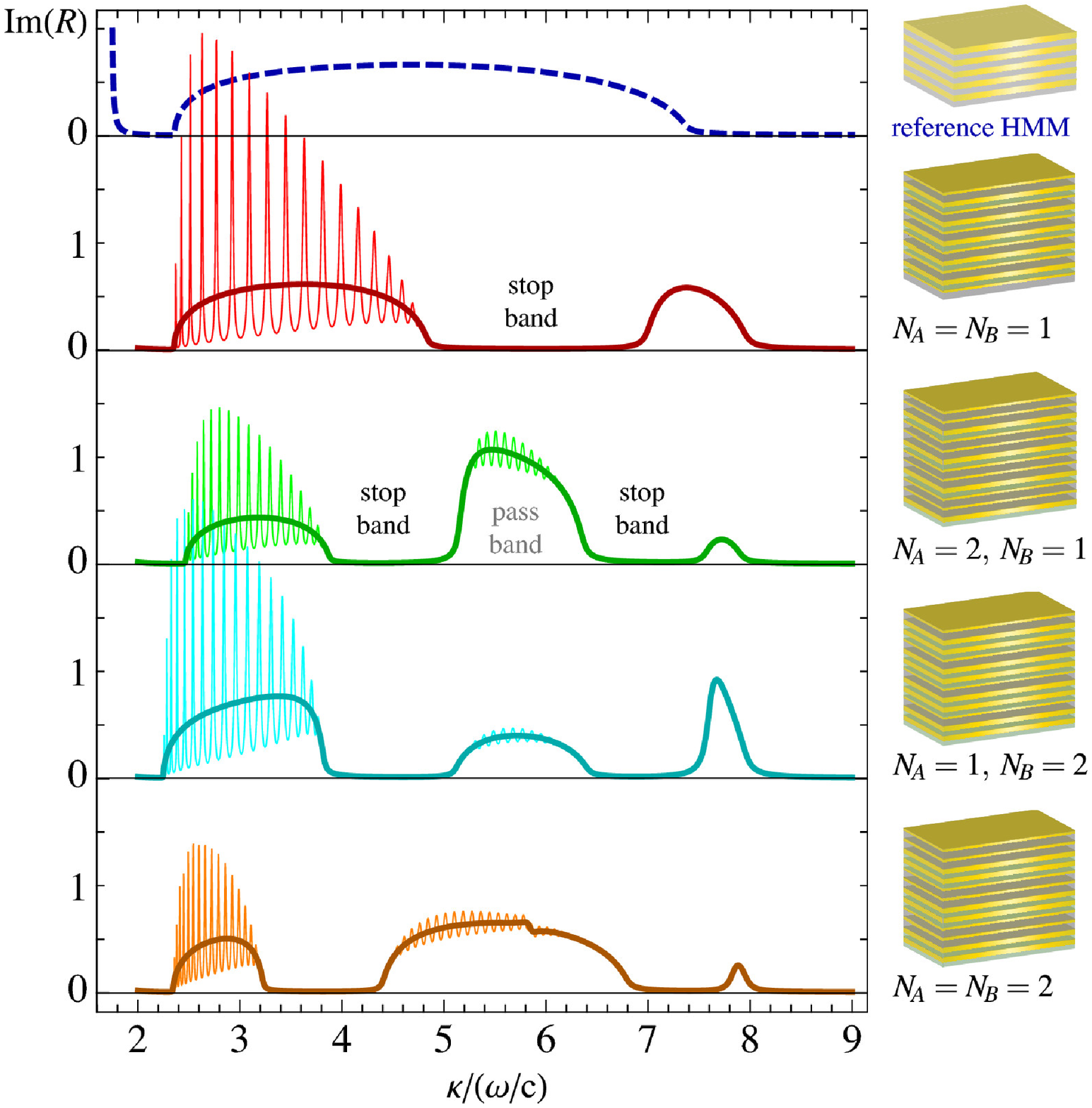}
\par\end{centering}

\caption{(Color online) Schematics and high-$k$ band characterization of the
structure with $N_{A}=N_{B}=1$ vs. structures with doubled number
of subperiods ($N_{A}=2,$ $N_{B}=1$; $N_{A}=1,$ $N_{B}=2$; $N_{A}=N_{B}=2$)
with $M=20$ superperiods. Other parameters are the same as in Fig.~\ref{FIG:mirror-realizable}.
Full metal losses are considered. \label{FIG:multigap}}
\end{figure}

As a final example, we briefly touch upon the possibility of band
gap engineering for VPP by using more complicated superstructure geometries
than simple periodicity. Note that doubling the number of subperiods
in some superlayers of thin-superlayer structures (changing from $N_{A,B}=1$
to $N_{A,B}=2$) drastically influences the corresponding layers:
they transform from quarter-wave-like to half-wave-like. This is expected
to result in band gap splitting. Indeed, Fig.~\ref{FIG:multigap}
shows that by simple alteration of the superstructure periodicity,
this multi-gap multiscale HMMs can be realized. Aside from providing
more versatility in the control over high-$k$ wave propagation, this
effect allows to push the band gap to the region with lower $\kappa$,
which are easier to excite and which are more prevalent in the emission
of a realistic source (finite-sized and/or located at a finite distance
from the HMM \cite{0ourOL11,0YuryPRA11}).

We can also envisage multiscale HMMs where superlayers lose their
periodicity entirely, while maintaining their long-range order, i.e.,
an ordered non-periodic superstructure. Out of the many examples of
such non-periodic geometries \cite{apMacia}, we will focus on a fractal
Cantor-like structure \cite{frHistSibilia,frHistSunJaggard,frOurPRE02}.
These structures are known for scalable and self-similar features
in their optical spectra closely related to their geometry \cite{frOurEPL04,frOurPNFA05,frOurPRE02}. 

\begin{figure}
\begin{centering}
\includegraphics[width=1\columnwidth]{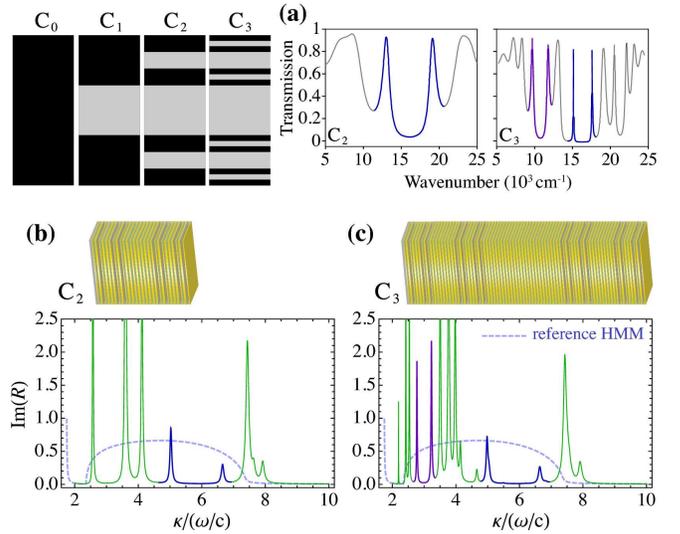}
\par\end{centering}

\caption{(Color online) (a) Schematics of middle-third Cantor section algorithm
and an example of the optical spectrum of a Cantor multilayer \cite{frOurPRE02,frOurEPL04}.
(b) Schematics and high-$k$ band characterization of a second-generation
Cantor structure ($C_{2}$, 9 superlayers). (c) Same as (b) but for
the third-generation Cantor structure ($C_{3}$, 27 superlayers).
The substructure geometry is the same as in Fig.~\ref{FIG:mirror-realizable}.
Full metal losses are considered. \label{FIG:fractal}}
\end{figure}

Specifically, we will consider the simplest middle-third Cantor sequence
(Fig.~\ref{FIG:fractal}a), described by the inflation rules \begin{equation}
\mathrm{A}\to\mathrm{AAA},\quad\mathrm{B}\to\mathrm{BAB},\label{eq:substitution}\end{equation}
applied to a single layer of the type B (the {}``seed'') several
times to form the Cantor structure of a given number of generation.
This procedure yields the following sequence: \begin{equation}
\begin{gathered}C_{0}=\mathrm{B},\; C_{1}=\mathrm{BAB},\; C_{2}=\mathrm{BABAAABAB},\\
C_{3}=\mathrm{BABAAABABAAAAAAAAABABAAABAB}.\end{gathered}
\label{eq:sequence}\end{equation}
It can also be written as a recurrent relation, \begin{equation}
C_{n+1}=C_{n}\left(\mathrm{A}\right)^{3^{n}}C_{n},\label{eq:sequence_recur}\end{equation}
which underlies its geometrical self-similarity and gives rise to
self-similar features in the optical spectra (Fig.~\ref{FIG:fractal}a)
\cite{frOurEPL04}. 

Using the substructure geometry similar to the previous cases as the
building blocks for A- and B-layers in Eqs.~\eqref{eq:substitution}--\eqref{eq:sequence_recur},
we analyze the fractal multiscale HMMs corresponding to the second-
and third-generation Cantor structures (containing 9 and 27 superlayers,
respectively). The results are shown in Fig.~\ref{FIG:fractal}b--c.
It can be seen that characteristic scalable signatures of the Cantor
spectra can be observed for VPPs in the high-$\kappa$ region of the
wave vector space. These features are distorted compared to the traditional
fractal multilayers because of the non-QW nature of the sublayers
involved. Nonetheless, they are observed in a realistic structure
with metal losses taken into account. So, we can expect that known
relations between geometrical and optical properties in deterministic
non-periodic multilayers should also be manifest in HMMs with corresponding
superstructure geometry. As a result, we can make use of the extensive
knowledge of PBG properties of multilayers \cite{frHistSibilia} to
be able to control VPP propagation in multiscale HMMs with significant
freedom.

\section{Conclusions\label{SEC:CONCLUSIONS}}

To summarize, we have demonstrated that a multiscale approach can
be used to control large-wavevector, bulk plasmonic waves (volume
plasmon polaritons) in multilayer metal-dielectric HMMs. As a proof
of concept, we have proposed the design of Bragg reflectors and Fabry-P{\'e}rot
resonators for these VPP waves. The designs consists of two levels
of structuring: (i) a substructure of subwavelength metal and dielectric
layers, responsible for creating hyperbolic dispersion, and (ii) a
superstructure, which constitutes wavelength-scale variation of metal
filling fraction and exerts PBG effects on VPPs. Band gaps and resonances
for VPPs have been demonstrated by examining the Fresnel reflection
coefficient in the large-wavevector region. More complicated superstructure
geometries such, as fractal Cantor-like multiscale HMMs, have also
been studied.

Along with proof-of-concept designs involving very large numbers of
layers, more realistic thin-superstructure designs have been proposed,
containing several tens of layers with thickness on the order of 10~nm,
which is within reach of modern fabrication technology. It has also
been shown that mechanisms to mitigate material absorption, for example
introducing gain-based compensation in dielectric layers \cite{0LoslessHMM},
make PBG effects more pronounced in all of the considered structures.

Our results show that VPPs can be directly controlled by the PBG effects,
which may be enabling for employing VPPs to transmit optical signals.
Using a great variety of superstructure geometries in the proposed
multiscale approach is promising in the design of HMM-based devices
with predetermined wavevector-space distribution of bulk plasmonic
waves. Such devices can be used in hyperlenses with tailored properties,
as well as to probe and tailor light-matter interaction phenomena
of nearby emitters (such as atoms and molecules) in the evanescent-wave
domain.
\begin{acknowledgments}
The authors wish to acknowledge inspiring discussions with K. Busch.
This work has received partial financial support from the People Programme
(Marie Curie Actions) of the European Union\textquoteright{}s 7th
Framework Programme FP7-PEOPLE-2011-IIF under REA grant agreement
No.~302009 (Project HyPHONE), as well as from the Natural Sciences
and Engineering Research Council of Canada (NSERC).
\end{acknowledgments}
\appendix

\section{Fresnel's reflection coefficients in the evanescent-wave domain\label{sec:APPENDIX}}

In this Appendix, we would like to discuss the physical meaning of
the Fresnel reflection coefficient at a plane interface between two
media for such values of the in-plane component of the wave vector,
$\kappa$, that the waves in one or both of the media can be evanescent.
The goal is to understand which attributes of the complex reflection
coefficient (real part, imaginary part, amplitude, and phase) have
direct physical interpretation, and to show that this interpretation
changes significantly when the incident wave is evanescent rather
than propagating. We need this knowledge in order to elucidate the
relation between the existence of propagating waves in the HMM structures
and the non-zero imaginary part of the reflection coefficients introduced
in the text in Eqs.~\eqref{eq:rHMM}, \eqref{eq:Airy}, \eqref{eq:Fresnel},
and \eqref{eq:Airy_bilayer}--\eqref{eq:Airy_inf}, as well as through
the transfer matrix calculation procedure.

We use the interface between two non-magnetic, isotropic media as
an example, and begin by recalling the expression for the reflection
coefficient for a $p$-polarized wave at such an interface, \begin{equation}
r_{ij}(\kappa)=\frac{w_{i}(\kappa)\epsilon_{j}-w_{j}(\kappa)\epsilon_{i}}{w_{i}(\kappa)\epsilon_{j}+w_{j}(\kappa)\epsilon_{i}},\label{eq:app_r_interface}\end{equation}
assuming that there are no losses; therefore, we assume that $\epsilon_{i,j}$
are purely real (positive or negative), and, in turn, that $w_{i,j}$,
defined in the same way as in Eq.~\eqref{eq:w}, are both one of
the following:\begin{equation}
w_{i}(\kappa)\text{ is }\begin{cases}
\text{real and positive,} & \epsilon_{i}>0\text{ and }\kappa<\tfrac{\omega}{c}\sqrt{\epsilon_{i}}\\
\text{positive imaginary,} & \text{otherwise}.\end{cases}\label{eq:app_w_conditions}\end{equation}
As mentioned in Section~\ref{SEC:PROPAGATION}, the first case corresponds
to a wave that is propagating in medium $i$, whereas the second case
corresponds to a wave that is evanescent in medium $i$. Depending
on which of the cases of Eq.~\eqref{eq:app_w_conditions} takes place
for the two media ($i$ and $j$) in Eq.~\eqref{eq:app_r_interface},
four scenarios can be identified: 
\begin{enumerate}
\item Both $w_{i}$ and $w_{j}$ are real, and hence $r_{ij}$ is real.
This is the usual Snell's refraction scenario: according to Eq.~\eqref{eq:app_w_conditions},
this is only possible if both $\epsilon_{i}$ and $\epsilon_{j}$
are non-negative so that the waves in both media are propagating,
resulting in $|r_{ij}|<1$, which means that the transmitted wave
carries some of the incident energy away. 
\item $w_{i}$ is real but $w_{j}$ is imaginary; the incident and reflected
waves are propagating but the transmitted wave is evanescent. This
is the total reflection scenario: either conventional total internal
reflection on a dielectric-dielectric interface beyond the critical
angle ($\sqrt{\epsilon_{j}}<\kappa c/\omega<\sqrt{\epsilon_{i}}$),
or total reflection from a dielectric-metal interface ($\epsilon_{j}<0<\epsilon_{i}$).
Here the reflection coefficient is of the form $(a-ib)/(a+ib)$ and
therefore $|r_{ij}|=1$, signifying that the transmitted evanescent
wave transfers no energy away from the interface.
\item Both $w_{i}$ and $w_{j}$ are imaginary. Similar to \#1 above, this
means that $r_{ij}$ must be real; however, $|r_{ij}|$ can be below
or above unity depending on the signs of $\epsilon_{i,j}$. This is
the scenario where all waves (incident, reflected, and transmitted)
are evanescent, and no energy transfer through the interface occurs.
If we were to change the medium $i$ so as to support a propagating
incident wave (e.g. by increasing its refractive index if it is a
dielectric), this scenario would change to \#2.
\item Finally, $w_{i}$ is imaginary but $w_{j}$ is real. Similar to \#2
above, $r_{ij}$ is of the form $(a-ib)/(a+ib)$ and therefore has
an imaginary part. This is the {}``reversed total reflection'' scenario
when the incident and reflected waves are evanescent but the transmitted
wave is\emph{ propagating}; if we were to change the medium $i$ so
as to support a propagating incident wave, this scenario would change
to \#1.
\end{enumerate}
Suppose, now, that we know nothing about the nature of medium $j$
beyond the interface and only treat that region as a {}``black box''
(or the {}``sample''). We still know, and can choose, the medium
$i$ in front of the interface (the {}``cladding''), and we can
perform some sort of ellipsometry-type reflectivity measurements on
the interface. An important observation about the four scenarios listed
above is that with the first two of them (\#1 and \#2), it is the
\emph{amplitude} of the reflection coefficient $|r_{ij}(\kappa)|$
that characterizes the sample's behavior at this particular $\kappa$;
we can conclude that $|r_{ij}|=1$ means that there are no propagating
waves in the sample (everything is totally reflected), and $|r_{ij}|<1$
means that there are propagating waves in the sample. On the other
hand, the \emph{phase} of the reflection coefficient (or individually
its real or imaginary part) cannot be attributed such physical significance.
Indeed, adding a cladding layer of thickness $d$ in front of the
sample will not change anything in the physical system but will, according
to Eqs.~\eqref{eq:Airy_bilayer} and Fig.~\ref{FIG:bilayers}b,
change $r_{ij}(\kappa)$ to $r'_{ij}(\kappa)=r{}_{ij}(\kappa)e^{2iw_{i}(\kappa)d}$,
making its phase dependent on $d$ and therefore, arbitrary. 

On the other hand, if we apply the same line of reasoning to scenarios
\#3 and \#4, we see that adding a similar cladding layer of thickness
$d$ in front of the sample, with similar transformation $r_{ij}(\kappa)\to r'_{ij}(\kappa)=r{}_{ij}(\kappa)e^{2iw_{i}(\kappa)d}$,
will change the \emph{amplitude} rather than the phase of the reflection
coefficient, since we are in the regime where $w_{i}$ is imaginary;
hence, it is $|r_{ij}|$ that can be made largely arbitrary. Therefore,
it is now the \emph{phase} of the reflection coefficient that characterizes
the physics of the sample: $\arg\, r_{ij}\ne0$ indicates that there
are bulk propagating waves in the sample while $\arg\, r_{ij}=0$
(real $r_{ij}$) unambiguously means that there are no such waves
(although surface waves at the interface may still exist). 

For practical purposes, whenever $r_{ij}$ is non-zero, we can introduce
a modified criterion based on the imaginary part of $r_{ij}$ rather
than on its phase. In these terms, $\mathrm{Im}\, r_{ij}=0$ (real
$r_{ij}$) signifies the absence of bulk propagating waves in the
sample, whereas $\mathrm{Im}\, r_{ij}\ne0$ indicates their presence,
as confirmed in Fig.~\ref{FIG:character}. We stress here that the
magnitude of $\mathrm{Im}\, r_{ij}$ still carries no direct physical
significance in terms of characterizing the sample since $|r_{ij}|$
can be arbitrary; it is only whether it is zero or non-zero that is
meaningful in the rigorous sense. However, we can relax our criterion
somewhat, saying that $\mathrm{Im}\, r_{ij}\approx0$ implies the
absence of propagating waves in the sample, and significantly non-zero
$\mathrm{Im}\, r_{ij}$ implies their presence, as is demonstrated
for the Bloch waves in Fig.~\ref{FIG:character}b. Vague as the words
{}``significantly non-zero'' are, the criterion in this form was
demonstrated to be useful in a broad range of parameters, including
complex multilayers and lossy structures, as confirmed by calculating
the field distribution at corresponding $\kappa$; the only regime
where we expect it to break down would be the case of high losses,
where any non-arbitrary distinction between propagating and evanescent
waves would be difficult.

We note, finally, that while the above analysis is carried out for
an interface between two isotropic media, it remains valid if the
sample is a homogenized HMM. Indeed, we see that Eq.~\eqref{eq:rHMM}
is essentially similar to Eq.~\eqref{eq:app_r_interface} as regards
the applicability of Eq.~\eqref{eq:app_w_conditions} and the subsequent
reasoning; the sole reason for the explicit use of Eq.~\eqref{eq:app_r_interface}
was to ease the explanation by being able to introduce a simple expression
for $w_{j}$ using $\epsilon_{j}$. Moreover, as outlined in Section~\ref{SEC:PROPAGATION}
in the discussion of Eq.~\eqref{eq:Airy_inf}, the reasoning remains
applicable to more complex samples such as infinite (and to some extent,
finite) multilayers.

\end{comment}
{}
\end{thebibliography}


\begin{thebibliography}{40}
\bibitem{0-PendryFocusIssue}J. Pendry, {}``Focus Issue: Negative
refraction and metamaterials -- Introduction,'' Opt. Express 11(7),
639 (2003).

\bibitem{1-Giant05}M. Kuwata-Gonokami, N. Saito, Y. Ino, M. Kauranen,
K. Jefimovs, T. Vallius, J. Turunen, and Y. Svirko, {}``Giant optical
activity in quasi-two-dimensional planar nanostructures,'' Phys.
Rev. Lett. 95(22), 227401 (2005).

\bibitem{2Smith03}D. R. Smith, D. Schurig, J. J. Mock, P. Kolinko,
and P. Rye, {}``Partial focusing of radiation by a slab of indefinite
media,'' Appl. Phys. Lett. 84(13), 2244--2246 (2004).

\bibitem{2aSmith04}D. R. Smith, D. Schurig, and P. Kolinko, {}``Negative
refraction in indefinite media,'' J. Opt. Soc. Am. B 21(5), 1032--1043
(2004).

\bibitem{2bSmith07}A. Degiron, D. R. Smith, J. J. Mock, B. J. Justice,
and J. Gollub, {}``Negative index and indefinite media waveguide
couplers,'' Appl. Phys. A 87 (2), 321--328 (2007).

\bibitem{2cSurfaceIndef}W. Yan, L. Shen, L. Ran, and J. A. Kong,
{}``Surface modes at the interfaces between isotropic media and indefinite
media,'' J. Soc. Am. A 24(2), 530--535 (2007).

\bibitem{0HyperlensOE06}Z. Jacob, L. V. Alekseyev, and E. Narimanov,
{}``Optical Hyperlens: Far-field imaging beyond the diffraction limit,''
Opt.~Express 14(18), 8247--8256 (2006). 

\bibitem{0NarimanovAPB10}Z. Jacob, J.-Y. Kim, G.V. Naik, A. Boltasseva,
E. E. Narimanov, and V. M. Shalaev, {}``Engineering photonic density
of states using metamaterials,'' Appl. Phys. B 100(1), 215--218 (2010).

\bibitem{0NarimanovAPL12}Z. Jacob, I. I. Smolyaninov, and E.E. Narimanov,
{}``Broadband Purcell effect: Radiative decay engineering with metamaterials,''
Appl. Phys. Lett. 100(18), 181105 (2012).

\bibitem{NarimanovStefanBoltzmann}C. Simovski, S. Maslovski, I. Nefedov,
and S. Tretyakov, ``Optimization of radiative heat transfer in hyperbolic
metamaterials for thermophotovoltaic applications,'' Opt. Express
21(12), 14988--15013 (2013).

\bibitem{0SmolyaninovJOSA}I. I. Smolyaninov and Yu-Ju Hung, {}``Modeling
of time with metamaterials,'' J.~Opt. Soc.~Am.~B 28(7), 1591--1595
(2011).

\bibitem{5NarimanovPRL10}I. I. Smolyaninov and E. E. Narimanov, {}``Metric
signature transitions in optical metamaterials,'' Phys. Rev. Lett.
105(6), 067402 (2010).

\bibitem{0NoginovAPL09}M. A. Noginov, Yu. A. Barnakov, G. Zhu, T.
Tumkur, H. Li, and E. E. Narimanov, {}``Bulk photonic metamaterial
with hyperbolic dispersion,'' Appl. Phys. Lett. 94(15), 151105 (2009).

\bibitem{3NoginovOL10}M. A. Noginov, H. Li, Yu. A. Barnakov, D. Dryden,
G. Nataraj, G. Zhu, C. E. Bonner, M. Mayy, Z. Jacob, and E. E. Narimanov,
{}``Controlling spontaneous emission with metamaterials,'' Opt.
Lett. 35(11), 1863--1865 (2010). 

\bibitem{6bSchillingAPL10}J. Kanungo and J. Schilling, {}``Experimental
determination of the principal dielectric functions in silver nanowire
metamaterials,'' Appl. Phys. Lett. 97(2), 021903 (2010).

\bibitem{ReviewJacob}C. L. Cortes, W. Newman, S. Molesky, and Z.
Jacob, ``Quantum nanophotonics using hyperbolic metamaterials,'' J.~Opt.
14, 063001 (2012).

\bibitem{ReviewKildishev}V. Drachev, V. A. Podolskiy, and A. V. Kildishev,
``Hyperbolic Metamaterials: new physics behind a classical problem,''
Opt. Express 21, 15048--15064 (2013).

\bibitem{0ourOL11}O. Kidwai, S. V. Zhukovsky, and J. E. Sipe, {}``Dipole
radiation near hyperbolic metamaterials: applicability of effective-medium
approximation,'' Opt. Lett. 36(13) , 2530--2532 (2011).

\bibitem{YuryOL11}I. Iorsh, A. Poddubny, A. Orlov, P. Belov, and
Yu. Kivshar, {}``Spontaneous emission enhancement in metal-dielectric
metamaterials,'' Phys. Lett. A 376(3), 185--187 (2012).

\bibitem{AvrPRB}I. Avrutsky, I. Salakhutdinov, J. Elser, and V. Podolskiy,
\textquotedblleft{}Highly confined optical modes in nanoscale metal-
dielectric multilayers,\textquotedblright{} Phys. Rev. B 75(24), 241402(R)
(2007).

\bibitem{LasPhotRev}S. Ishii, A. V. Kildishev, E. Narimanov, V. M.
Shalaev, and V. P. Drachev, \textquotedblleft{}Sub-wavelength interference
pattern from volume plasmon polaritons in a hyperbolic medium,\textquotedblright{}
Laser Photon. Rev. 7(2), 365--271 (2013).

\bibitem{blFengOE05}S. Feng, J. M. Elson, and P. L. Overfelt, {}``Optical
properties of multilayer metal-dielectric nanofilms with all-evanescent
modes,'' Opt. Express 13(11), 4113--4124 (2005).

\bibitem{blOrenOE11}G. Rosenblatt and M. Orenstein, {}``Competing
coupled gaps and slabs for plasmonic metamaterial analysis,'' Opt.
Express 19(21), 20372--20385 (2011).

\bibitem{Schilling}J. Schilling, \textquotedblleft{}Uniaxial metallo-dielectric
metamaterials with scalar positive permeability,\textquotedblright{}
Phys. Rev. E 74(4), 046618 (2006).

\bibitem{AvrAPL}J. Elser, V. A. Podolskiy, I. Salakhutdinov, and
I. Avrutsky, ``Nonlocal effects in effective-medium response of nanolayered
metamaterials,'' Appl. Phys. Lett. 90(19), 191109 (2007).

\bibitem{ourFocusOE13}S. V. Zhukovsky, O. Kidwai, and J. E. Sipe,
\textquotedblleft{}Physical nature of volume plasmon polaritons in
hyperbolic metamaterials,\textquotedblright{} Opt. Express 21, 14982
(2013).

\bibitem{0Saarinen}J. J. Saarinen and J. E. Sipe, {}``A Green function
approach to surface optics in anisotropic media,'' J. Mod. Opt. 55(1),
13--32 (2008). 

\bibitem{0ourHMMPRA}O. Kidwai, S. V. Zhukovsky, and J. E. Sipe, {}``Effective-medium
approach to planar multilayer hyperbolic metamaterials: Strengths
and limitations,'' Phys. Rev. A 85(5), 053842 (2012).

\bibitem{BookYarivYeh}A. Yariv and P. Yeh, \emph{Optical Waves in
Crystals} (New York: Wiley, 1983).

\bibitem{0LoslessHMM}X. Ni, S. Ishii, M. Thoreson, V. Shalaev, S.
Han, S. Lee, and A. Kildishev, ''Loss-compensated and active hyperbolic
metamaterials,'' Opt. Express 19(25), 25242--25254 (2011).

\bibitem{ourPRA10}S. V. Zhukovsky, {}``Perfect transmission and
highly asymmetric light localization in photonic multilayers,'' Phys.
Rev. A 81(5), 053808 (2010).

\bibitem{ourOC13}S. V. Zhukovsky, L. G. Helt, D. Kang, P. Abolghasem,
A. S. Helmy, and J. E. Sipe, \textquotedblleft{}Analytical description
of photonic waveguides with multilayer claddings: Towards on-chip
generation of entangled photons and Bell states,\textquotedblright{}
Opt. Commun. 301\textendash{}302, 127\textendash{}140 (2013).

\bibitem{BetterPlasmonics}P. R. West, S. Ishii, G. V. Naik, N. K.
Emani, V. M. Shalaev, and A. Boltasseva, {}``Searching  for  better
 plasmonic   materials,'' Laser Photon. Rev. 4(6), 795--808 (2010).

\bibitem{0YuryPRA11}A. N. Poddubny, P. A. Belov, and Yu. S. Kivshar,
{}``Spontaneous radiation of a finite-size dipole emitter in hyperbolic
media,'' Phys. Rev. A 84(2), 023807 (2011).

\bibitem{apMacia} E. Maci\'{a}, {}``The role of aperiodic order
in science and technology,'' Rep. Prog. Phys. 69(2), 397--441 (2006).

\bibitem{frHistSunJaggard}X. Sun and D. Jaggard, {}``Wave interactions
with generalized Cantor bar fractal multilayers,'' J. Appl. Phys.
70(5), 2500--2507 (1991).

\bibitem{frHistSibilia}C. Sibilia, I. S. Nefedov, M. Scalora, and
M. Bertolotti, {}``Electromagnetic mode density for finite quasi-periodic
structures,'' J. Opt. Soc. Am. B 15(7), 1947--1952 (1998).

\bibitem{frOurPRE02}A. V. Lavrinenko, S. V. Zhukovsky, K. S. Sandomirskii,
and S. V. Gaponenko, {}``Propagation of classical waves in nonperiodic
media: Scaling properties of an optical Cantor filter,'' Phys. Rev.
E 65(3), 036621 (2002).

\bibitem{frOurEPL04}S. V. Zhukovsky, A. V. Lavrinenko, and S. V.
Gaponenko, \textquotedblleft{}Spectral scalability as a result of
geometrical self-similarity in fractal multilayers,\textquotedblright{}
Europhys. Lett. 66(3), 455--461 (2004).

\bibitem{frOurPNFA05}S. V. Zhukovsky and A. V. Lavrinenko, \textquotedblleft{}Spectral
self-similarity in fractal one-dimensional photonic structures,\textquotedblright{}
Photonics and Nanostructures \textendash{} Fundamentals and Applications
3(2--3), 129--133 (2005).%
\begin{comment}
\begin{thebibliography}{40}
\bibitem{key-2}-------------------

\bibitem{7NiAPB11}X. Ni, G. Naik, A. Kildishev, Yu. Barnakov, A.
Boltasseva, and V. Shalaev, Appl. Phys. B 103, 553 (2011)

\bibitem{0JacobReview}C. L. Cortes, W. Newman, S. Molesky, and Z.
Jacob, J.~Opt. 14, 063001 (2012).

\bibitem{0Podolskiy}I. Avrutsky, I. Salakhutdinov, J. Elser, and
V. Podolskiy, Phys. Rev. B 75, 241402(R) (2007).

\bibitem{3xNarimanovPre09}Z. Jacob, I. I. Smolyaninov, and E. E.
Narimanov, http://arxiv.org/abs/0910.3981.

\bibitem{6xSoukoulisPRB09}A. Fang, T. Koschny, and C. M. Soukoulis,
Phys. Rev. B 79, 245127 (2009).

\bibitem{RafalJOSA11}R. Koty\'{n}ski, T. J. Antosiewicz, K. Kr\'{o}l,
and K. Panajotov, J. Opt. Soc. Am. A 28, 111 (2011).

\bibitem{RafalAPA11}R. Koty\'{n}ski, T. Stefaniuk, and A. Pastuszczak,
Appl. Phys. A 103, 905 (2011).

\bibitem{RafalReview}R. Koty\'{n}ski, Opto-Electronics Rev. 18,
366 (2010).

\bibitem{7Sipe81}J. E. Sipe, Surf. Sci. 105, 489 (1981).

\bibitem{BornWolf}M. Born and E. Wolf, \emph{Principles of Optics},
6th ed. (Cambridge Univ. Press, 1997)

\bibitem{9SipePRA85}J. M. Wylie and J. E. Sipe, Phys. Rev. A 32,
2030 (1985).

\bibitem{GoldRefIndex}P. G. Etchegoin, E. C. Le Ru, and M. Meyer,
J. Chem. Phys. 125, 164705 (2006).

\bibitem{AluminaIndex}T. S. Eriksson, A. Hjortsberg, G. A. Niklasson,
and C. G. Granqvist, Appl. Opt. 20, 2742 (1981).

\bibitem{GraphiteAPL11}J. Sun, Ji Zhou, Bo Li, and F. Kang, Appl.
Phys. Lett. 98, 101901 (2011).

\bibitem{BeriniReview}P. Berini, Adv. Opt. Photon. 1, 484 (2009). 

\bibitem{SipeOE05}J. Saarinen, S. Weiss, P. Fauchet, and J. E. Sipe,
Opt. Express 13, 3754 (2005).

\bibitem{TransMetal}M. J. Bloemer and M. Scalora, Appl. Phys. Lett.
72, 1676 (1998).

\bibitem{6NarimanovCLEO10}E. Narimanov, M. A. Noginov, H. Li, and
Y. Barnakov, in QELS Conference: OSA Technical Digest (CD) (Optical
Society of America, 2010), paper QPDA6.
\end{thebibliography}
\end{document}